\documentclass[a4paper]{article}
\usepackage{dx14}
\usepackage[T1]{fontenc}
\usepackage{ae,aecompl}
\usepackage{amsfonts}
\usepackage{amsmath}
\usepackage{amsthm}
\usepackage{graphicx}

\usepackage{times}

\title{An Overview of Direct Diagnosis and Repair Techniques\\in the \textsc{WeeVis} Recommendation Environment\thanks{This  work has been conducted in the  project \textsc{PeopleViews} funded by the Austrian Research Promotion Agency (843492).}}
\author%
{%
Alexander Felfernig$^1$ \and Stefan Reiterer$^1$ \and Martin Stettinger$^1$ \and Michael Jeran$^1$\\
$^1$Graz University of Technology, Graz, Austria\\
e-mail: \{alexander.felfernig, stefan.reiterer, martin.stettinger, mjeran\}@ist.tugraz.at\\
}

\begin{document}

\maketitle

\begin{abstract}
Constraint-based recommenders support users in the identification of items (products) fitting their wishes and needs. Example domains are financial services and electronic equipment. In this paper we show how divide-and-conquer based (direct) diagnosis algorithms (no conflict detection is needed) can be exploited in  constraint-based recommendation scenarios. In this context, we provide an overview of the MediaWiki-based recommendation environment \textsc{WeeVis}.
\end{abstract}

\section{Introduction}

Constraint-based recommenders \cite{FelfernigGula2006,felfernigburke08} support the identification of relevant items from large and often complex assortments. Example item domains are electronic equipment  \cite{Felfernig2006} and financial services \cite{felfernigvita2007}. In contrast to collaborative filtering \cite{Konstan1997} and content-based filtering \cite{Pazzani1997}, constraint-based recommendation relies on an explicit representation of recommendation knowledge. Two  major  types of knowledge sources are exploited for the definition of a constraint-based recommendation task  \cite{felfernigburke08}. First, knowledge about the given set of customer requirements. Second, recommendation knowledge that is represented as a set of items and a set of constraints that help to establish a relationship between  requirements and the item assortment.

Diagnosis techniques can be useful in the following situations: (1) in situations where it is not possible to find a solution for a given set of user (customer) requirements, i.e., the requirements are inconsistent with the recommendation knowledge base and the user is in the need for repair proposals to find a way out from the "no solution could be found" dilemma; (2) if a recommendation knowledge base is inconsistent with a set of test cases that has been defined for the purpose of regression testing, the knowledge engineer needs support in figuring out the responsible faulty constraints. 

For situation (1) we sketch how model-based diagnosis \cite{Reiter1987} can be applied for the identification of faulty constraints in a given set of customer requirements. In this context  efficient divide-and-conquer based algorithms can be applied to the diagnosis and repair of inconsistent requirements. In a similar fashion, such algorithms can be applied for the diagnosis of inconsistent recommender knowledge bases (the knowledge base itself can be inconsistent, or alternatively, inconsistencies can be induced by  test cases used for regression testing).

The diagnosis approaches presented in this paper have been integrated into  \textsc{WeeVis} which is a MediaWiki-based recommendation environment for complex products and services. In the line of the Wikipedia\footnote{www.wikipedia.org.}  idea to support communities of users in the cooperative development of Web content, \textsc{WeeVis} is an environment that supports all the functionalities available for the creation of Wiki pages. Additionally, it allows the inclusion of constraint-based recommender applications that help to work up existing knowledge and present this in a compressed and intuitive fashion.

The contributions of this paper are the following. First, we sketch how efficient divide-and-conquer based algorithms can be applied for solving diagnosis and repair tasks in constraint-based recommendation scenarios. Second, we sketch how diagnosis and repair approaches can be integrated into Wiki technologies\footnote{www.mediawiki.org.} and with this be made accessible to a large user group. Third, we discuss challenges for future research that have to be tackled to advance the state-of-the-art in constraint-based recommendation.

The remainder of this paper is organized as follows. In Section \ref{workingexample} we discuss properties of constraint-based recommendation tasks. Thereafter, we introduce an example recommendation knowledge base. In Section \ref{inconsistentrequirements} we show how divide-and-conquer based algorithms can be applied for the diagnosis and repair of inconsistent requirements. Thereafter we show how such algorithms can be applied to the identification of faulty constraints in  knowledge bases (see Section \ref{inconsistentknowledgebase}). Related and future work are discussed in Section \ref{relatedfuturework}. We conclude the paper with Section \ref{conclusions}.

\section{Working Example}\label{workingexample}

\begin{figure*}[ht!]
	\centering
	\fbox{
		\includegraphics[width=0.80\textwidth]{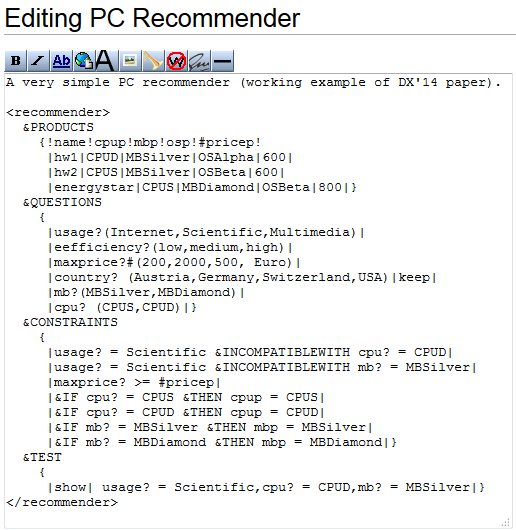}
	}
	\caption{Example \textsc{WeeVis} PC Recommender definition  (MediaWiki "Edit" mode).}
	\label{recommenderdefinition}
\end{figure*}

 In the remainder of this paper we will use \emph{personal computer recommendation} as working example. Roughly speaking, a recommendation task consists of selecting those items that match the user requirements. In the context of personal computers, the recommender user has to specify his/her requirements regarding, for example, the intended usage, the maximum accepted price, and the cpu type. Since \textsc{WeeVis} is a MediaWiki-based environment, the definition of a recommender knowledge base is supported in a textual fashion (see Figure \ref{recommenderdefinition}).

On the basis of a set of requirements, the recommender system determines alternative solutions (the consideration set) and presents these to the user. If no solution could be found for the  given requirements, repair alternatives are determined which support users in getting out of the "no solution could be found" dilemma (see Figure \ref{recommenderexecution}).

Constraint-based recommendation requires the explicit definition of questions (representing alternatives for user requirements), properties of the items, and constraints. An example of a recommendation knowledge base is shown in Figure \ref{recommenderdefinition}. The \textsc{WeeVis} tag \&QUESTIONS enumerates variables that describe  user requirements where \emph{usage} specifies the intended use of the computer, \emph{eefficiency} represents the required energy efficiency, \emph{maxprice} denotes the upper price limit specified by the user, \emph{country} represents the country of the user, \emph{mb} represents the type of motherboard, and \emph{cpu} the requested central processing unit. If a variable is associated with a \emph{keep} tag, this variable is not taken into account in the diagnosis process. For example,  \emph{country?} is associated with a \emph{keep} tag; for this reason, it will not be part of any diagnosis presented to the recommender user. Other examples of such attributes are a person's age and gender.

In addition to variables representing potential user requirements, a recommendation knowledge base includes the definition of variables that represent item properties (represented by the \textsc{WeeVis} tag \&PRODUCTS). In our example, cpu$_p$ represents the CPU included in the item, mb$_p$ specifies the included motherboard, os$_p$ represents the installed operating system, and price$_p$ is the overall price. Furthermore, the set of items (products) must be specified that can be recommended to users. A simplified item assortment is included in Figure \ref{recommenderdefinition} as part of the item properties. Our example assortment of items consists of the entries \emph{hw1}, \emph{hw2}, and \emph{energystar}.

Incompatibility constraints describe combinations of requirements that lead to an inconsistency. The description related to the \textsc{WeeVis} tag \&CONSTRAINTS includes an incompatibility relationship between the variable \emph{usage} and the variable \emph{cpu}. For example, computers with a CPUD must not be sold to users interested in scientific calculations.

Filter constraints describe the relationship between user requirements and items. A simple example of such a filter constraint is \emph{maxprice} $\ge$ \emph{price}$_p$, i.e., the price of an recommended item must be equal or below the maximum accepted price specified by the customer (see the \textsc{WeeVis} tag \&CONSTRAINTS in Figure \ref{recommenderdefinition}).

Finally, \textsc{WeeVis} supports the definition of test cases (see also Section \ref{inconsistentknowledgebase}) which can be used to specify the intended behavior of a recommender knowledge base (\textsc{WeeVis} tag \&TEST). After changes to the knowledge base, regression tests can be triggered on the basis of the defined test suite. The |show| tag specifies whether the recommender system user interface should show the status of the test case (satisfied or not) -- see, for example, Figure \ref{kbdiag}.

On a formal level, a recommendation knowledge base can be represented as a constraint satisfaction problem \cite{Mackworth1977} with two sets of variables V = U $\cup$ P and the corresponding constraints C = COMP $\cup$ PROD $\cup$ FILT. In this context, $u_i \in U$ are variables describing possible user requirements (e.g., \emph{usage} or \emph{maxprice}) and $p_i \in P$ are variables describing item (product) properties (e.g., \emph{mb}$_p$ or \emph{price}$_p$). 

The recommendation knowledge base specified in Figure \ref{recommenderdefinition} can be transformed into a constraint satisfaction problem where \&QUESTIONS represents $U$, \&PRODUCTS represents $P$ and $PROD$, and \&CONSTRAINTS represents $COMP$ and $FILT$.\footnote{PROD is assumed to be represented as a single constraint in disjunctive normal form where each conjunct is an item.} Given such a recommendation knowledge base we are able to determine concrete recommendations on the basis of a specified set of user (customer) requirements. Requirements collected are represented in terms of constraints, i.e., R = \{$r_1, r_2, ..., r_k$\} represents a set of user requirements. 

After having identified the set of alternative solutions (recommended items or consideration set), this result is presented to the user. In constraint-based recommendation scenarios, the ranking of items is often performed on the basis of Multi-Attribute Utility Theory (MAUT) where items are evaluated on the basis of a given set of interest dimensions. For further details on the ranking of items in constraint-based recommendation scenarios we refer to \cite{Felfernig2013}.

\section{Diagnosis and Repair of Requirements}\label{inconsistentrequirements}

In situations where the given set of requirements $r_i \in R$ (unary constraints defined on variables of U such as \emph{maxprice} $\leq$ \emph{500}) become inconsistent with the recommendation knowledge base (C), we are interested in repair proposals that indicate for a subset of these requirements change operations with a high probability of being accepted by the user.  On a more formal level we now introduce a definition of a customer requirements diagnosis task and a corresponding diagnosis (see Definition 1).

\emph{Definition 1 (Requirements Diagnosis Task)}. Given a set of requirements R and a set of constraints C (the recommendation knowledge base), the diagnosis task it to identify a minimal set $\Delta$ of constraints (the diagnosis) that have to be removed from R such that R -  $\Delta$ $\cup$ C is consistent.

An example of a set of requirements for which no solution can be identified is R = \{$r_1$: \emph{usage} = \emph{Scientific}, $r_2$ :\emph{eefficiency} = \emph{high}, $r_3$: \emph{maxprice} = \emph{1700}, $r_4$: \emph{country} = \emph{Austria}, $r_5$:\emph{mb} = \emph{MBSilver}, $r_6$: \emph{cpu} = \emph{CPUD}\}. The recommendation knowledge base induces two minimal conflict sets (CS) \cite{junker04quickxplain} in R which are $CS_1$: \{$r_1, r_6$\} and $CS_2$: \{$r_1,r_5$\}. For these conflict sets we have two alternative diagnoses which are  $\Delta_1$:\{$r_5, r_6$\} and $\Delta_2$:\{$r_1$\}. The pragmatics, for example, of $\Delta_1$ is that at least $r_5$ and $r_6$ have to be adapted in order to be able to find a solution. How to determine such diagnoses on the basis of a HSDAG (hitting set directed acyclic graph)  is shown, for example, in \cite{felfernig2004}. 

Approaches based on the construction of hitting sets typically rely on conflict detection  \cite{junker04quickxplain,felfernig2004}. In interactive settings, where only preferred diagnoses (leading diagnoses) should be presented, hitting set based approaches tend to become too inefficient since conflict sets have to be determined before a diagnosis can be presented \cite{felfernig2012,dx2010}. This was the major motivation for the development of the \textsc{FastDiag} algorithm \cite{felfernig2012,dx2010,FelfernigBenavidesetal2013}, which is a divide-and-conquer based algorithm that enables the determination of minimal diagnoses without the need of conflict determination and HSDAG construction. This way of determining minimal diagnoses can also be denoted as \emph{direct diagnosis} since no conflict set determination is needed in this context.

\begin{figure}[ht!]
	\centering
	\fbox{
		\includegraphics[width=0.27\textwidth]{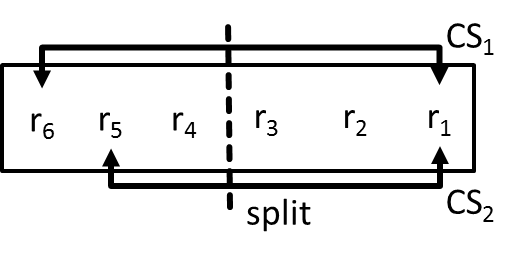}
	}
	\caption{Divide-and-conquer principle of \textsc{FastDiag} ($CS_1$ and $CS_2$ are assumed to be conflict sets). The set of requirements $R$ = \{$r_1, ..., r_6$\} is split in the middle. If a diagnosis is already contained in the first part of the split ($R$ - \{$r_6, r_5, r_4$\} is consistent), there is no need to further investigate the right part for further diagnosis elements. This way, half of the potential diagnosis elements can be eliminated in one step (consistency check).}
	\label{fastdiagprinciple}
\end{figure}

\textsc{FastDiag} can be seen as an inverse \textsc{QuickXPlain} \cite{junker04quickxplain} type algorithm which relies on the following basic principle (see Figure \ref{fastdiagprinciple}). Given, for example, a set R = \{$r_6, r_5, ..., r_1$\} and a diagnosis (see Definition 1) is contained in \{$r_6,r_5,r_4$\} (first part of the split), then there is no need of further evaluating  \{$r_3,r_2,r_1$\}, i.e., the latter set is consistent. The similarity to \textsc{QuickXPlain} is the following. If a minimal conflict is contained in  \{$r_6,r_5,r_4$\} there is no need to further search for conflicts in \{$r_3,r_2,r_1$\} since the algorithm determines one minimal conflict set at a time. Both algorithms (\textsc{FastDiag} and \textsc{QuickXPlain}) rely on a total lexicographical ordering \cite{junker04quickxplain,felfernig2012} which allows the determination of preferred minimal diagnoses (minimal conflict sets).

A minimal (preferred) diagnosis $\Delta$ can be used as a basis for the determination of corresponding repair actions, i.e., concrete measures to change user requirements in R in a fashion such that the resulting R' is consistent with C.

\emph{Definition 2 (Repair Task)}. Given a set of requirements R = \{$r_1,r_2,...,r_k$\} inconsistent with the constraints in C and a corresponding diagnosis $\Delta \subseteq R$ ($\Delta$ = \{$r_l, ..., r_o$\}), the corresponding repair task is to determine an adaption A = \{$r_l$', ..., $ r_o$'\} such that R - $\Delta$ $\cup$ A is consistent with C.

\begin{figure*}[ht!]
	\centering
	\fbox{
		\includegraphics[width=0.85\textwidth]{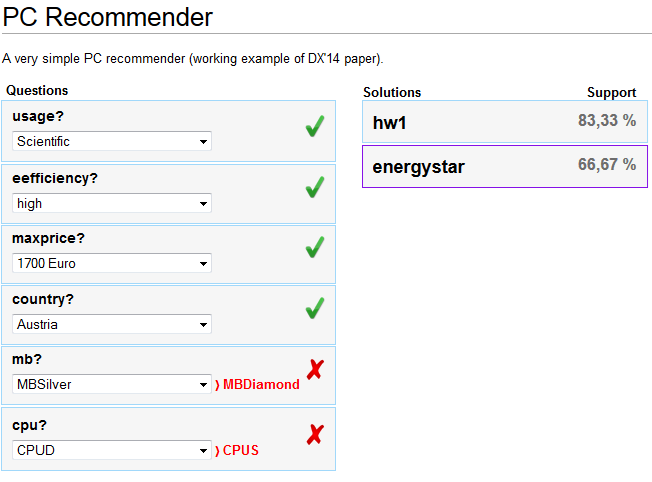}
	}
	\caption{PC recommender UI (MediaWiki "Read" mode). If the user selects the item \emph{energystar} on the right-hand side, a diagnosis with corresponding repair actions is depicted on the left-hand side.}
	\label{recommenderexecution}
\end{figure*}

In \textsc{WeeVis}, repair actions are determined conform to Definition 2. For each diagnosis $\Delta$ determined by \textsc{FastDiag} (currently, the first n=3 leading diagnoses are determined -- for details see \cite{felfernig2012}), the corresponding solution search  for R - $\Delta$ $\cup$ C returns a set of alternative repair actions (represented as adaptation A). In the following, all products that satisfy  R - $\Delta$ $\cup$ A are shown to the user (see the right hand side of Figure \ref{recommenderexecution}).

In the current \textsc{WeeVis} implementation, the total lexicographical ordering is derived from the order in which a user has entered his/her requirements. For example, if $r_1$: \emph{usage} = \emph{Scientific} has been entered before $r_5$: \emph{mb} = \emph{MBSilver} and $r_6$: \emph{cpu} = \emph{CPUD} then the underlying assumption is that $r_5$ and $r_6$ are of lower importance for the user and thus have a higher probability of being part of a diagnosis. In our working example $\Delta_1$ = \{$r_5, r_6$\}. The corresponding set of repair actions (solutions for R-$\Delta_1$ $\cup$ C) is A = \{$r_5$':mb=MBDiamond, $r_6$':cpu=CPUS\}, i.e., \{$r_1, r_2, r_3, r_4, r_5, r_6$\} - \{$r_5, r_6$\} $\cup$ \{$r_5$', $r_6$'\} is consistent. The item that satisfies R - $\Delta_1$ $\cup$ A is \{hw1\} (see the first entry in Figure \ref{recommenderexecution}). In a similar fashion, repair actions are determined for $\Delta_2$ - the recommended item is \{energystar\}.  The identified items (p) are finally ranked according to their support value (see Formula \ref{itemranking}).

\begin{equation}\label{itemranking}
support(p) = \frac{\# repair~actions~in~R'}{\#~requirements~in~R}
\end{equation}

\section{Knowledge Base Diagnosis}\label{inconsistentknowledgebase}
Recommendation knowledge is often subject to change operations. Due to frequent changes it is important to support quality assurance of recommendation knowledge. \textsc{WeeVis} supports the definition and execution of test cases\footnote{\textsc{WeeVis} supports the definition of positive test cases (test cases that should be consistent with the knowledge base).} which define the intended behavior of the recommender knowledge base. If some test cases become inconsistent with a new version of the knowledge base, the causes of the unintended behavior must be identified. On a formal level a recommendation knowledge base (RKB) diagnosis task can be defined as follows (see Definition 3).

\emph{Definition 3 (RKB Diagnosis Task)}. Given a set C (the recommendation knowledge base) and a set T = \{$t_1, t_2, ..., t_q$\} of test cases $t_i$, the corresponding diagnosis task is it to identify a minimal set $\Delta$ of constraints (the diagnosis) that have to be removed from C such that $\forall t_i \in T$: $C - \Delta \cup t_i$ is consistent. 

An example  test case  which induces an inconsistency with the constraints in C is  $t$: \emph{usage} = \emph{Scientific} and \emph{cpu} = \emph{CPUD} and \emph{mb} = \emph{MBSilver} (see Figure \ref{recommenderdefinition}). $t$ induces two conflicts in the recommendation knowledge base which are $CS_1$: $\neg$(\emph{usage} = \emph{Scientific} $\land$ \emph{cpu} = \emph{CPUD}) and $CS_2$: $\neg$(\emph{usage} = \emph{Scientific} $\land$ \emph{mb} = \emph{MBSilver}). In order to make C consistent with $t$, both incompatibility constraints have to be deleted from C, i.e., are part of the diagnosis $\Delta$.

\begin{figure*}[ht!]
	\centering
	\fbox{
		\includegraphics[width=0.60\textwidth]{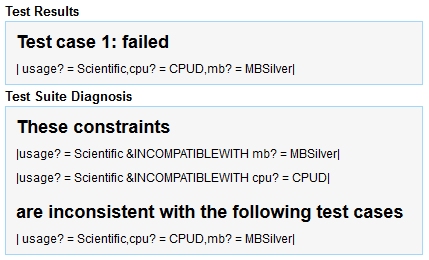}
	}
	\caption{PC recommender knowledge base: result of the diagnosis process presented in \textsc{WeeVis}.}
	\label{kbdiag}
\end{figure*}

Similar to the diagnosis of inconsistent requirements, the hitting set based determination of diagnoses for inconsistent knowledge bases is shown in \cite{felfernig2004}. This approach relies on the construction of a HSDAG determined on the basis of minimal conflict sets provided by conflict detection algorithm such as \textsc{QuickXPlain}. Diagnoses are determined in a breadth-first fashion, i.e., \emph{minimal cardinality} diagnoses of faulty constraints in C are returned first. 

In contrast to \cite{felfernig2004}, \textsc{WeeVis} includes a \textsc{FastDiag} based approach to knowledge base debugging that can also be applied in interactive settings. In this case, diagnoses are searched in C. In the case of requirements diagnosis, the total ordering of the requirements is related to user preferences (in \textsc{WeeVis} derived from the instantiation order of variables). Total orderings of constraints in the context of knowledge base diagnosis are determined using criteria different from the diagnosis of inconsistent requirements, for example, age of constraints, frequency of quality assurance, and structural constraint complexity (see \cite{felfernigconfws2013}). An example screenshot of the \textsc{WeeVis} diagnosis presentation is depicted in Figure \ref{kbdiag}.

\section{Related and Future Work}\label{relatedfuturework}

\emph{Diagnosing Inconsistent Requirements}. Junker \cite{junker04quickxplain} introduced the \textsc{QuickXPlain} algorithm which is a divide-and-conquer based approach to the determination of minimal conflict sets (one conflict set at a time). Combining \textsc{QuickXPlain} with the hitting set directed acyclic graph (HSDAG) algorithm \cite{Reiter1987} allows for the calculation of the complete set of minimal conflicts. O'Sullivan et al. \cite{sullivan2007} show how to determine representative explanations (diagnoses) which fulfill the requirement that minimal subsets $\Delta_S$ of the complete set of diagnoses $\Delta_C$ should be determined that fulfill the criteria that if a constraint $c_i$ is contained in a diagnosis of  $\Delta_C$ it must also be part of at least one diagnosis in $\Delta_S$. Felfernig et al. \cite{felfernigmairitschmandl2009,felfernigijcai2013} show how to integrate similarity metrics, utility-, and probability-based approaches to the determination of leading diagnoses on the basis HSDAG-based search.

Felfernig and Schubert \cite{dx2010} introduce \textsc{FlexDiag} which is a top-down version of \textsc{FastDiag} allowing a kind of anytime diagnosis due to the fact that diagnosis granularity (size of constraints regarded as one component in the diagnosis process) can be parametrized. Felfernig et al. \cite{felfernig2012,dx2010} introduce the \textsc{FastDiag} algorithm that allows for a more efficient determination of diagnoses due to the fact the there is no need for determining conflict sets (= \emph{direct diagnosis}). \textsc{FastDiag} is a \textsc{QuickXPlain} style algorithm that follows a divide-and-conquer approach for the determination of minimal diagnoses. Note that in contrast to traditional HSDAG based approaches, \textsc{FastDiag} does not focus on the determination of minimal cardinality but preferred minimal diagnoses. A major issue for future work will be the development of diagnosis algorithms that are capable of performing intra-constraint debugging an thus help to better focus on the sources of inconsistencies. \textsc{FastDiag} is not restricted to the application in knowledge-based recommendation scenarios but generally applicable in consistency-based settings \cite{KnowledgeRepresentation}. For example, the same principles can  be applied in  knowledge-based configuration \cite{Stumptner,Sabin1998,Felfernigetal2014}. Further approaches to the determination of diagnoses for inconsistent knowledge bases can be found, for example, in \cite{McAreavey2014,MarquesSilva2014,MarquesSilv2013,Malitsky2014,Walter2013}.

\emph{Knowledge Base Maintenance}. The application of model-based diagnosis for the debugging of inconsistent constraint sets was first presented in \cite{paperBakker1993}. Felfernig et al. \cite{felfernig2004} show how to exploit test cases for the induction of conflict sets in knowledge bases which are then resolved on the basis of a hitting set based approach. In the line of the work of \cite{felfernig2012,dx2010} the performance of knowledge debugging can be improved on the basis of \textsc{FastDiag}. A detailed evaluation of the performance gains of \textsc{FastDiag} in the context of knowledge base debugging is within the focus of our future work. A detailed comparison between the performance of \textsc{FastDiag} and conflict-driven diagnosis of inconsistent requirements can be found, for example, in \cite{felfernig2012}.

Identifying redundant constraints is an additional issue in the context of knowledge base development and maintenance. Redundant constraints can deteriorate runtime performance and also be the cause of additional overheads in development and maintenance operations \cite{FelfernigDXCoreDiag2011}. Redundancy detection can be based on \textsc{QuickXPlain} especially in the case of an increasing number of redundant constraints. For a detailed discussion of alternative algorithms for redundancy detection in knowledge bases we refer to \cite{FelfernigDXCoreDiag2011}. A major focus of our future research will be the development of an intra-constraint redundancy detection, i.e., it will be possible to identify redundant subexpressions.

\section{Conclusions}\label{conclusions}

In this paper we provide an overview of the \textsc{WeeVis} environment with a special focus on the integrated diagnosis support. Diagnosis techniques integrated in \textsc{WeeVis} are the result of research in model-based diagnosis with a special focus on divide-and-conquer based (direct) algorithms that make diagnosis search more efficient in the case that  leading diagnoses are required. \textsc{WeeVis} is a publicly available MediaWiki-based environment for developing and maintaining constraint-based recommender applications.

\bibliographystyle{unsrt}
\bibliography{dx14}

\end{document}